\documentclass[a4paper,11pt]{article}

\usepackage{jheppub} 

\usepackage{graphicx}
\usepackage{amsmath,amssymb,mathrsfs}
\usepackage{epsf}
\usepackage{epsfig}
\usepackage[font=small,skip=0pt,justification=raggedright]{caption}

\newcounter{fig}   \newcommand{\lbfig}[1]{\refstepcounter{fig}
\label{#1} }

\newcommand{\bea}{\begin{eqnarray}}
\newcommand{\eea}{\end{eqnarray}}
\newcommand{\be}{\begin{equation}}
\newcommand{\ee}{\end{equation}}

\newcommand{\re}[1]{(\ref{#1})}

\newcommand{\eqn}{\begin{eqnarray}}
\newcommand{\eqnx}{\end{eqnarray}}

\def\({\left(}
\def\){\right)}
\def\vp{\varphi}
\def\br{\begin{eqnarray}}
\def\er{\end{eqnarray}}

\begin{document}

\title{Exact Self-Dual Skyrmions}

\author{L.A. Ferreira$^{\dagger}$ and
Ya.~Shnir$^{\ddagger\star\mathsection}
$}
\affiliation{$^{\dagger}$Instituto de F\'{i}sica de S\~{a}o Carlos; IFSC/USP;
Universidade de  S\~{a}o Paulo, USP;
Caixa Postal 369, CEP 13560-970, S\~{a}o Carlos-SP, Brazil\\
$^{\ddagger}$BLTP, JINR, Dubna 141980, Moscow Region, Russia\\
$^{\star}$Department of Theoretical Physics and Astrophysics, BSU, Minsk 220004, Belarus\\
$^{\mathsection}$Department of Theoretical Physics, Tomsk State Pedagogical University, Russia
}

\abstract{We introduce a Skyrme type model with the target space being  the sphere $S^3$ and with an action possessing, as usual, quadratic and quartic terms in field derivatives. The novel character of the model is that  the strength of the couplings of those two terms are allowed to depend upon the space-time coordinates. The model should therefore be  interpreted  as an effective theory, such that those couplings correspond in fact to low ener\-gy expectation values of fields belonging to a more fundamental theory at high energies. The theory possesses a self-dual sector that saturates the Bogomolny bound leading to an energy depending linearly on the topological charge. The self-duality equations are conformally invariant in three space dimensions leading to a toroidal ansatz and exact self-dual Skyrmion solutions. Those solutions are labelled by two integers and, despite their toroidal  character,  the energy density  is spherically symmetric when those integers are equal and oblate or prolate otherwise. }

\maketitle

\section{Introduction}
%
Self-dual field configurations possess very nice physical and mathematical properties, and they are important in  the study of non-linear aspects of field theories possessing topological solitons.  The best known examples are
the instanton solutions of the Yang-Mills theory in four dimensional Euclidean space \cite{Belavin:1975fg} and the
self-dual Bogomol'nyi-Prasad-Sommerfield (BPS) monopoles in the 3+1 dimensional Yang-Mills-Higgs theory \cite{BPS-mono,BPS-mono2}.
The self-dual solitons satisfy first order differential equations which yields the absolute minimum of the energy, and  by construction they are
also solutions of the full dynamical system of the field equations. Another feature of the self-dual field configurations is that the corresponding
topological solitons always saturate the topological bound, their static energy
(or the Euclidean action in the case of the Yang-Mills instantons) depends linearly on the topological charge. Moreover, there are
very elegant mathematical methods of construction of various multi-soliton configurations in these models, the Nahm equation \cite{Nahm:1979yw}
and the algebraic Atiyah-Hitchin-Drinfeld-Manin scheme \cite{Atiyah:1978ri}.

However the usual Skyrme model \cite{skyrme1},\cite{skyrme2}, which can be suggested as an effective low-energy theory of pions, do not
support self-dual equations \cite{Manton:1986pz},
the mass of the soliton solutions for this model, the Skyrmions, is always above the topological lower bound in a given topological sector
\cite{Faddeev:1976pg}, even though in compact spaces it is possible to saturate a bound \cite{canfora,Ferreira:2013bia}. As a consequence, there is no exact mathematical scheme of construction of multi-soliton solutions of the Skyrme model,
the only way to obtain these solutions
in any topological sector, is to implement various numerical methods, some of them are rather sophisticated, they usually need
a large amount of computational power.

Recently some modification of the Skyrme model was proposed to construct the soliton solutions which satisfy the first-order Bogomol'nyi-type equation
\cite{Adam:2010fg,Adam:2010ds,Sutcliffe:2010et}. In the first case the conventional Skyrme  model was drastically changed via replacement of the usual
sigma model term and the quartic Skyrme term
with a term sextic in first derivatives and a potential \cite{Adam:2010fg,Adam:2010ds}. In the second case the
usual Skyrme model is coupled to the infinite tower of vector mesons \cite{Sutcliffe:2010et}.
These self-dual models are directly related to the usual Skyrme model since they can be considered as submodels of a general model of that type.
Further, it was shown very recently that
the standard Skyrme model without the potential term can be expressed as a sum of two BPS submodels with different solutions \cite{Adam:2017pdh}.
The corresponding submodels, however, are not directly related to the generalized Skyrme model of any type.

Another modification of the Skyrme model, which supports self-dual solutions and has an exact BPS bound,
was suggested in \cite{Ferreira:2013bia}. Similar to the
usual Skyrme model, or its generalizations, the field of the new model is a map from compactified coordinate
space $S^3$ to the $SU(2)$ group space. The corresponding first order equations are equivalent to the so-called force free equation
well known in solar and plasma physics, see e.g. \cite{marsh}. The drawback of this construction is that, due to an argument by Chandrasekhar \cite{chandra},  these equations
does not possess finite energy solutions on $\mathbb{R}^3$, although it supports regular solutions on three sphere $S^3$ \cite{Ferreira:2013bia}.

In this paper we propose generalization of the self-dual Skyrme-type model discussed in \cite{Ferreira:2013bia},
which possess the regular solution on $\mathbb{R}^3$.
Similar to the usual Skyrme model, we consider a nonlinear scalar  sigma-model, parameterized by two complex scalar fields $Z_a$, $a=1,2$,
satisfying the constraint $Z_a^*\,Z_a=1$. The action of the model is given by
\be
S= \int d^4x\,\left( \frac{m^2(x^{\rho})}{2}  A_{\mu}^2  - \frac{1}{4  e^2(x^{\rho})} H_{\mu\nu}^2\right)
\label{ua845}
\ee
with two couplings $m(x^{\rho})$ and $e(x^{\rho})$, dependent upon the space-time coordinates, which are of dimension of mass and
dimensionless, respectively.
In addition, $\mu\, ,\,\nu=0,1,2,3$, and
\be
A_{\mu}= \frac{i}{2}\left(Z_a^*\partial_{\mu} Z_a-Z_a\partial_{\mu} Z_a^*\right) \qquad\qquad {\rm and} \qquad\qquad
H_{\mu\nu}=\partial_{\mu} A_{\nu}-\partial_{\nu} A_{\mu}.
\label{adef}
\ee
The model \re{ua845} is similar to the one considered in \cite{Ferreira:2013bia}, with the main difference being the fact that the coupling constants now are allowed to depend upon the space-time coordinates. That plays a crucial role in the properties of the model. In the first place, it circumvents the famous Chandrasekhar's  argument \cite{chandra} that prevents the existence of finite energy solutions extending over the whole $\mathbb{R}^3$ space.   In addition, as we explain below, it renders  the self-duality equations  conformally invariant in the three dimensional space $\mathbb{R}^3$. As it is usual in many effective field theories, coupling constants that depend upon the space-time coordinates correspond in fact to low energy expectation values of fields belonging to a more fundamental theory at higher energies. At the end of the paper we shall discuss some possibilities for the introduction of a dilation type field that could account for the space-time dependent coupling constants appearing in  \re{ua845}.  We now discuss the properties of the self-dual sector of the theory  \re{ua845}.

\section{The self-duality equations}
It will be convenient for our purposes to represent the corresponding static energy functional via the dual of $H_{ij}$, defined as
\be
B_i=\frac{1}{2} \, \varepsilon_{ijk}\, H_{jk}\, , \qquad \qquad  i,j,k=1,2,3
\label{dualh}
\ee
Then we can write the static energy associated to \re{ua845} as
\be
E= \frac{1}{2}\,\int d^3x\, \left( m^2\left({\vec r}\right)\,A_{i}^2  + \frac{1}{ e^2\left({\vec r}\right)}\, B_i^2\right)
\label{energyua845}
\ee
In order to have finite energy solutions the fields   $Z_a$, $a=1,2$, have to approach fixed constant values at  spatial infinity, and so as long as topological arguments are concerned,  we
can compactify  the physical space  $\mathbb{R}^3$ to $S^3$. Thus the field of the  model \re{ua845} becomes a map
$Z_a: S^3 \to S^3$. The mapping is labeled by the topological invariant $Q=\pi_3(S^3)$, which is the  winding number of the field configuration, and it can be calculated by the following integral
\be
Q=\frac{1}{12\,\pi^2}\,\int d^3x\, \varepsilon_{abcd}\,\varepsilon_{ijk}\,\Phi_{a}\,\partial_i\Phi_{b}\,\partial_j\Phi_{c}\,\partial_k \Phi_{d}=
\frac{1}{4\,\pi^2}\,\int d^3 x\, A_i\,B_i,
\label{topcharge}
\ee
where we have written $Z_1\equiv \Phi_1+i\,\Phi_2$, $Z_2\equiv \Phi_3+i\,\Phi_4$, and $a,b,c,d=1,2,3,4$.

Note that on the right hand side of \re{topcharge} we have written $Q$ in terms of the vectors $A_i$ and $B_i$ defined in \re{adef} and \re{dualh},
respectively. Evidently, this structure reminds the Hopf invariant used in the theories with the target space being $S^2$,
like in the  Skyrme-Faddeev model \cite{Faddeev-Hopf}.
However, our target space is still $S^3$ and we are not projecting the map down to $S^2$ as is the case of the first Hopf map.

Next we follow the arguments presented in \cite{selfdual}.
Let us denote by $\chi_{\alpha}$, $\alpha=1,2,3$, the independent fields of the target space $S^3$.
The topological charge $Q$ given in  \re{topcharge} is invariant under infinitesimal smooth
(homotopic) deformations of the fields $\delta \chi_{\alpha}$, and so, without the use of the equations of motion,
one finds that $\delta Q=0$. Since the variations are arbitrary one gets from \re{topcharge} that
the vectors $A_i$ and $B_i$ have to satisfy
\be
B_i\frac{\delta A_i}{\delta \chi_{\alpha}}
-  \partial_{j}\left(B_i\frac{\delta A_i}{\delta \partial_{j}\chi_{\alpha}}\right)
+A_i\frac{\delta B_i}{\delta \chi_{\alpha}}
-  \partial_{j}\left(A_i\frac{\delta B_i}{\delta \partial_{j}\chi_{\alpha}}\right)=0 .
\label{identity}
\ee
On the other hand, the static Euler-Lagrange equations associated to \re{ua845} are given by
\be
m^2\left({\vec r}\right)\, A_i\frac{\delta A_i}{\delta \chi_{\alpha}}
-  \partial_{j}\left(m^2\left({\vec r}\right)\, A_i\frac{\delta A_i}{\delta \partial_{j}\chi_{\alpha}}\right)
+\frac{1}{e^2\left({\vec r}\right)}\,B_i\frac{\delta B_i}{\delta \chi_{\alpha}}
-  \partial_{j}\left(\frac{1}{e^2\left({\vec r}\right)}\,B_i\frac{\delta B_i}{\delta \partial_{j}\chi_{\alpha}}\right)=0 .
\label{euler}
\ee
If one now imposes the self-duality equation
\be
m({\vec r})\, A_i=\pm\, \frac{1}{e({\vec r})}\, B_i
\label{selfdualeq}
\ee
one gets that \re{identity} becomes
\be
\begin{split}
\pm& m({\vec r})\, e({\vec r})\, A_i\frac{\delta A_i}{\delta \chi_{\alpha}}
-  \partial_{j}\left(\pm m({\vec r})\,e({\vec r})\, A_i\frac{\delta A_i}{\delta \partial_{j}\chi_{\alpha}}\right)\\
\pm& \frac{1}{m({\vec r})\,e({\vec r})}\,B_i\frac{\delta B_i}{\delta \chi_{\alpha}}
-  \partial_{j}\left(\pm \frac{1}{m({\vec r})\,e({\vec r})}\,B_i\frac{\delta B_i}{\delta \partial_{j}\chi_{\alpha}}\right)=0 .
\end{split}
\label{identity2}
\ee
If, in addition, we  impose that
\be
m({\vec r})=m_0\,f({\vec r})\qquad\qquad\qquad e({\vec r})=e_0\,f({\vec r})
\label{nicecoupling}
\ee
with $m_0$ and $e_0$ being constants, then \re{identity2} becomes the same as \re{euler}.
The conclusion is that, for the choice \re{nicecoupling}, the self-duality equation \re{selfdualeq} implies
\re{euler}, when the identity \re{identity}, coming from the topological charge, is used.
Therefore, using \re{nicecoupling}, the static energy \re{energyua845} becomes
\be
E= \frac{1}{2}\,\int d^3x \left[\left(m_0^2\, f^2\, A_i^2+ \frac{1}{e_0^2\, f^2}\, B_i^2\right)\right]
\label{energyua845-f}
\ee
which can be written as
\be
E = \frac12 \int d^3 x\,\left( m_0 f A_{i}  \mp \frac{1}{e_0 f}\, B_i\right)^2
\pm\,\frac{m_0}{e_0}\,\int d^3x\, A_{i}  \, B_i
\ee
Therefore the lower energy bound is
\be
E \geq 4\,\pi^2\, \frac{m_0}{e_0}\,\mid Q\mid
\label{eng-bound}
\ee
which is saturated for the solutions of the self-duality equation
\be
m_0 e_0 \, f^2\, A_i =  \pm  B_i
\label{sd-eqs}
\ee
For such self-dual field configurations we have
\be
E= m_0^2\,\int d^3x\, f^2 A_{i}^2= \frac{1}{e_0^2}\,\int d^3x\, \frac{B_{i}^2}{f^2}
\label{energyselfdualf}
\ee
Note that if we treat $f$ as an independent field then the static Euler-Lagrange equation coming from \re{energyua845-f} is given by
\be
m_0^2\, f^2\, A_i^2= \frac{1}{e_0^2\, f^2}\, B_i^2
\ee
which certainly follows from the self-duality equation \re{sd-eqs}. Therefore, the first order self-duality equations \re{sd-eqs} imply all the static second order Euler-Lagrange equations associated to the theory \re{ua845}, when the coupling constants have the form given in \re{nicecoupling}.
As we shall see, the dilaton function $f(r)$ can regularize solutions of the self-duality equation
providing a way to evade the usual arguments \cite{chandra,Ferreira:2013bia} that there can be no finite energy solutions of the force free equation.

\section{Conformal symmetry of the model}
Remarkably, the the self-dual sector of the  model \re{ua845} is invariant under conformal transformations in three space dimensions. In order to see it, we will follow the approach of
\cite{babelon} and consider a general
infinitesimal space transformations of the form
$\delta x_i= \zeta_i $, such that
\be
\delta Z_a=0;\qquad\qquad
\delta \partial_i Z_a=-\partial_i\zeta_j\, \partial_j Z_a \, .
\label{conf-trans}
\ee
Therefore
\be
\delta A_i= -\partial_i\zeta_j\,A_j\qquad\qquad\qquad
\delta H_{ij}=-\partial_i\zeta_k H_{kj}-\partial_j\zeta_k H_{ik} \, ,
\ee
and
\be
\begin{split}
\delta B_i= -\varepsilon_{ijk} \,\partial_j\zeta_l H_{lk}=-\partial_j\zeta_l \varepsilon_{ijk} \,\varepsilon_{lkm} \, B_m
=\partial_j\zeta_i\, B_j- \partial_j\zeta_j\, B_i
\end{split}
\ee
Let us consider how the self-duality equations \re{sd-eqs} change under such transformations. It is convenient to  write them in the form
\be
\Lambda_i\equiv \lambda\, f^2\,A_i-B_i=0\qquad\qquad \qquad \lambda=\eta\, m_0\,e_0\qquad\qquad \eta=\pm 1 \, ,
\label{lambdadef}
\ee
and so
\be
\begin{split}
\delta \Lambda_i=& 2\,\frac{\delta f}{f}\, \lambda\, f^2\,A_i-\partial_i\zeta_j\,\lambda\,f^2\,A_j
- \partial_j\zeta_i\, B_j+ \partial_j\zeta_j\, B_i \\
=&\left[2\,\frac{\delta f}{f}\,\delta_{ij}-(\partial_i\zeta_j+\partial_j\zeta_i)+\partial_l\zeta_l\,\delta_{ij}
\right]\,\lambda\,f^2\,A_{j} \, .
\end{split}
\ee
Hence, in order to remain invariant with respect to the transformations \re{conf-trans}, the variations of the space coordinates $\zeta_i$
must satisfy
\be
\partial_i\zeta_j+\partial_j\zeta_i=2\,D\,\delta_{ij}
\label{conformal}
\ee
for some function $D$. Therefore,
\be
\delta \Lambda_i=\left[2\,\frac{\delta f}{f}+D\,\right]\,\lambda\,f^2\,A_{i}
\ee
and the self-duality equation \re{sd-eqs} remains invariant if
\be
\delta f=-\frac{D}{2}\,f
\label{conf-f}
\ee
As is was shown in \cite{babelon}, the transformations satisfying \re{conformal} are actually the
conformal transformations. Indeed, we have that the possibilities are
\bea
\begin{array}{lll}
 \zeta_i^{(P_j)}=\varepsilon^{(P_j)}\, \delta_{ij} & D^{(P_j)}=0
& {\rm (translations)}\\
 \zeta^{R_{jk}}_i= \varepsilon^{(R_{jk})}\, (\delta_{ki}\, x_j-\delta_{ji}\, x_k)\qquad j\neq k \qquad & D^{(R_{ij})}=0 & {\rm (rotations)}\\
 \zeta^{(d)}_i=\varepsilon^{(d)}\, x_i & D^{(d)}=\varepsilon^{(d)}& {\rm (dilatations)}\\
 \zeta^{(c_j)}_i=\varepsilon^{(c_j)}\,\left(x_i\,x_j-\frac{1}{2}\,x_l^2\,\delta_{ij}\right)
 & D^{(c_j)}=\varepsilon^{(c_j)}\, x_j \quad & \mbox{\rm (special conf.)}
 \end{array}
 \label{conftransdef}
\eea
Therefore the self-duality equations \re{sd-eqs} are invariant under conformal transformations in three dimensional space.
Note that $f$ is a scalar field under translations and rotations but not under
dilatations and special conformal transformations. Further, one can check that
\be
\delta A_i^2= - 2\,D\, A_i^2\;;
\qquad\qquad
\delta B_i^2= - 4\,D\, B_i^2\;;
\qquad\qquad
\delta (A_i\,B_i)= - 3\,D\, A_i\, B_i
\label{ababtransf}
\ee
and the volume element transforms as
\be
\delta (d^3x)=3\, D\, d^3x
\ee
Hence both the static energy functional \re{energyua845-f} and the topological charge \re{topcharge} are
conformally invariant.

\section{The toroidal ansatz and exact Skyrmion solutions on $\mathbb{R}^3$}

Here we again follow the reasonings of \cite{babelon} to construct an ansatz for our self-duality equations, which  is invariant under the diagonal subgroup of two commuting $U(1)$'s in the conformal group and other two commuting $U(1)$'s in the internal symmetry group of the model \re{ua845}.
Note that the model \re{ua845}
is invariant under the $U(2)$ global transformations
\be
\left(\begin{array}{c}
Z_1\\
Z_2
\end{array}\right)\rightarrow U\,
\left(\begin{array}{c}
Z_1\\
Z_2
\end{array}\right)\qquad \qquad\qquad U \in U(2)
\ee
The Cartan  subgroup of the $U(2)$ includes two commuting $U(1)$ elements, namely
\be
Z_1\rightarrow e^{i\alpha}\,Z_1\qquad\qquad \qquad Z_2\rightarrow Z_2
\label{firstu1}
\ee
and
\be
Z_1\rightarrow Z_1\qquad\qquad \qquad Z_2\rightarrow e^{i\beta}\,Z_2
\label{secondu1}
\ee
In addition we also have, in the conformal group in three dimensions,
two commuting $U(1)$ elements, which correspond to the vector fields
${\cal V}_{\zeta}= \zeta_i\,\partial_i$ with (see \cite{babelon})
\bea
\partial_{\varphi}&\equiv& {\cal V}_{\varphi}= x_2\partial_1-x_1\partial_2
\label{dphidef}\\
\partial_{\xi}&\equiv& {\cal V}_{\xi} =\frac{x_3}{a}\left(x_1\partial_1+x_2\partial_2\right)+\frac{1}{2\,a}\left(a^2+x_3^2-x_1^2-x_2^2\right)\partial_3
\label{dxidef}
\eea
with $a$ being a length scale factor, and where we have introduced two angles, $\varphi$ and $\xi$, such that the vectors fields,
${\cal V}_{\varphi}$ and ${\cal V}_{\xi}$, ge\-ne\-ra\-te  rotations  along those angular directions. Note, that  ${\cal V}_{\varphi}$ is the generator of rotations on the plane $x_1\,\,x_2$. On the other hand,
${\cal V}_{\xi}$ is a linear combination of the special conformal generator $V^{(c_3)}=x_3\,x_i\partial_i-\frac{1}{2}\,x_l^2\,\partial_3$, and the translation generator $V^{(P_3)}= \partial_3$ (see \re{conftransdef}). One can easily check that indeed $\left[\partial_{\varphi}\,,\, \partial_{\xi}\right]=0$. The third curvilinear coordinate in $\mathbb{R}^3$ which is orthogonal to $\varphi$ and $\xi$ is
\be
z=\frac{4\,a^2(x_1^2+x_2^2)}{(x_1^2+x_2^2+x_3^2+a^2)^2}
\label{zinv}
\ee
One can check that indeed $\partial_{\varphi} z=\partial_{\xi} z=0$. It turns out that
$(z\,,\,\xi\,,\,\varphi)$ constitute the toroidal coordinates in $\mathbb{R}^3$ defined as\footnote{We have replaced the usual toroidal coordinate $\eta$ by $z$, these coordinates are related as $z=\tanh^2 \eta$, with $\eta>0$.}
\be
x^1= \frac{a}{p}\, \sqrt{z}\,\cos \varphi\;;\qquad\quad
x^2= \frac{a}{p}\, \sqrt{z}\,\sin \varphi\;;\qquad\quad
x^3= \frac{a}{p}\, \sqrt{1-z}\,\sin \xi
\label{toroidal}
\ee
where
\be
p=1-\sqrt{1-z}\,\cos\xi\qquad\qquad\qquad 0\leq z\leq 1\qquad\qquad 0\leq \varphi\, ,\,\xi\leq 2\,\pi
\label{pdef}
\ee
We now want field configurations that are invariant under the diagonal subgroup of the tensor product of the internal $U(1)$ defined in \re{firstu1} and the external $U(1)$ generated by $\partial_{\varphi}$ given in \re{dphidef}. In addition we want those same field configurations to be invariant under the diagonal subgroup of the tensor product of the internal $U(1)$ defined in \re{secondu1} and the external $U(1)$ generated by $\partial_{\xi}$ given in \re{dxidef}. That brings us to the toroidal ansatz defined by
\be
Z_1= \sqrt{F(z)}\, e^{i\, n\,\varphi}\qquad\qquad\qquad
Z_2= \sqrt{1-F(z)}\, e^{i\,m\,\xi}
\label{parameterizez}
\ee
where $m$ and $n$ are two integers, to keep the configuration single valued in $\mathbb{R}^3$.

In order to proceed it is convenient to write  the self-duality equation  in terms of vector calculus notation, and so we have that \re{sd-eqs}  can be written as
\be
{\vec \nabla}\wedge {\vec A} = \eta\, m_0\, e_0\, f^2({\vec r})\, {\vec A};\qquad\qquad\qquad
 \eta=\pm 1 \, .
 \label{vectorselfdual}
\ee
Writing ${\vec A}$ in terms of the unit vectors of the toroidal coordinates as ${\vec A}=\frac{V_{z}}{h_z} \, {\vec e}_{z}+\frac{V_{\xi}}{h_{\xi}} \, {\vec e}_{\xi}+\frac{V_{\varphi}}{h_{\varphi}} \, {\vec e}_{\varphi}$ (see \re{unitvectors}), we have that $V_{\zeta}=\frac{i}{2}\(Z_a^*\partial_{\zeta} Z_a-Z_a \partial_{\zeta} Z_a^*\)$, with $\zeta\equiv z , \xi, \varphi$, and where the scaling factors $h_{\zeta}$ are defined in \re{scalingfactors}. Therefore, \re{vectorselfdual} can be written  in components as
\be
\begin{split}
\kappa\, f^2\, V_{z} & = \frac{p}{2\,z\left(1-z\right)}\, \partial_{\xi}V_{\varphi}\\
\kappa\, f^2\, V_{\xi}& = -2\,(1-z)\,p\,\partial_{z}V_{\varphi}\\
\kappa\, f^2\, V_{\varphi}& =2\, z\, p\, \left[\partial_{z} V_{\xi}-\partial_{\xi}V_{z}\right]
\end{split}
\label{eqforv}
\ee
where we have introduced the  dimensionless quantity $\kappa\equiv \eta\,m_0\,e_0\,a$, with $\eta=\pm 1$ (see \re{lambdadef}). Substituting the ansatz \re{parameterizez} in the self-duality equations \re{eqforv} we get
\be
\begin{split}
\partial_{\xi}F &= 0\\
\frac{\kappa\,m\, f^2}{p}\left(1-F\right) &=-2\,\left(1-z\right)\,n\,\partial_{z}F\\
\frac{\kappa\,n\, f^2}{p} F&=-2\,z\,m\,\partial_{z}F
\label{eqforf}
\end{split}
\ee
Now we can eliminate the derivative $\partial_{z}F$ from this system, it yields
a simple algebraic solution of the self-duality equations \re{eqforf} for any values of the integers $m$ and $n$
\be
F=\frac{m^2\,z}{m^2\,z +n^2(1-z)} \qquad\qquad\qquad\qquad  f^2= \frac{2\,p}{m_0\,e_0\,a}~  \frac{\mid m\,n\mid}{[m^2\,z +n^2(1-z)]}
\label{solution-nm}
\ee
where, to keep $f$ real, we had to choose the sign of $\(m\, n\)$ to be related to the sign $\eta$ of the self-duality as $\eta=-{\rm sign}\(m\,n\)$.
Thus, the explicit form of the solution for the self-dual model \re{ua845} on $\mathbb{R}^3$ is
\be
\begin{split}
Z_1&= \sqrt{\frac{m^2\,z}{m^2\,z +n^2\left(1-z\right)}}\;\;\; e^{i\,n\,\varphi}\\
Z_2&= \sqrt{\frac{n^2\left(1-z\right)}{m^2\,z +n^2\left(1-z\right)}}\;\;\; e^{i\, m\,\xi}\\
f&= \sqrt{ \frac{2\,\mid m\,n\mid \left(1-\sqrt{1-z}\,\cos\xi\right)}{m_0\,e_0\,a\left[m^2\,z +n^2(1-z)\right]}}
\label{solution-fields}
\end{split}
\ee
The vector field ${\vec A}$ takes the following form when evaluated on the solutions \re{solution-fields}
\be
{\vec A}=- m\,n \,\frac{p/a}{m^2\,z+n^2\(1-z\)}\left[ {\vec e}_{\xi}\, n\, \sqrt{1-z}+{\vec e}_{\vp}\, m\, \sqrt{z}\right]
\label{A-vect}
\ee
and so
\be
A_i^2=m^2\,n^2 \,\frac{p^2/a^2}{m^2\,z+n^2\(1-z\)}
\ee
Note that ${\vec A}$ is tangent to the toroidal surfaces defined by $z={\rm constant}$. On the circle on the $x_1\,x_2$ plane defined by $z=1$ (see the appendix \ref{sec:appendix}), one has that
${\vec A}_{{\rm circle}}=-n\,{\vec e}_{\vp}/a$. At spatial infinity, where $z=0$ and $\xi=0$, one has that ${\vec A}_{{\rm infinity}}= 0$. On the $x_3$-axis, where $z=0$, one has ${\vec A}_{x_3-{\rm axis}}=-(m/a) (1-\cos\xi)\, {\vec e}_{\xi}$. 

Evaluating the static energy \re{energyselfdualf} on the solutions \re{solution-fields}, we get
\be
E= m_0^2\,\int d^3x\,  f^2\, {\vec A}^2
=4\,\pi^2\,\frac{m_0}{e_0}\mid m\,n\mid\, m^2\,n^2\,\int_0^1 dz\,
\frac{ 1} { \left[ m^2\,z + n^2(1-z) \right]^2}
\ee
Using the fact that
$$
\int_0^1dz\,\frac{1}{\left[m^2\,z +n^2(1-z)\right]^2}=\frac{1}{m^2\,n^2}
$$
one gets
\be
E=4\,\pi^2\,\frac{m_0}{e_0}\mid m\,n\mid
\ee
Further, using \re{sd-eqs} into the definition of the topological charge \re{topcharge}, we get that the solutions \re{solution-fields} have topological charges given by
\be
Q = \frac{1}{4\,\pi^2}\,\int d^3 x\, A_i\,B_i =  \frac{\eta\,m_0\,e_0}{4\,\pi^2}\,\int d^3 x\,f^2 A_i^2
=-m\,n
\ee
where we have used that the sign $\eta=\pm 1$, in the self-duality equation  \re{sd-eqs} is related to $m\,n$ as $\eta=-{\rm sign}\(m\,n\)$ (see below \re{solution-nm}).
Thus, these field configurations exactly saturate the topological bound \re{eng-bound} for any values of $n,m$.

Note that the solutions \re{solution-fields} are very similar to those exact solutions constructed in \cite{hopfion99}, and possessing in fact the same topological charges. The  model in \cite{hopfion99} however, is defined on target space $S^2$ and it does not possess a self-dual sector, even though it presents conformal symmetry in three space dimensions.

If one of the integers, $m$ or $n$ vanishes, the solutions \re{solution-fields} become trivial with $f=0$ everywhere in space. Note that in the particular case where $n=\pm m$ the general solution \re{solution-fields} is
reduced to
\be
Z_1= \sqrt{z}\,\,  e^{i\,n\,\varphi},\qquad
Z_2= \sqrt{1-z}\,\,  e^{\pm i\,n\,\xi},\qquad
f=\sqrt{\frac{2}{m_0e_0a}~ (1-\sqrt{(1-z)}\cos \xi)}
\label{reduced-sol}
\ee
Clearly, the field $Z_a$ in \re{reduced-sol} resembles the form of the solution of the model \re{ua845} on $S^3$, constructed in \cite{Ferreira:2013bia},
however the angular variables in the
latter case are related with the angular coordinates on the three-sphere and the function $f$, which is a regulator on $\mathbb{R}^3$, does not appear there.

Remind that in the toroidal coordinates \re{toroidal}
the spacial infinity corresponds to $z=0$, $\xi = 0$ and the origin corresponds to $z=0$, $\xi=\pi$. Evidently, for the solutions
\re{solution-fields} we have the asymptotic behavior
\be
Z_1(r\to \infty) = 0,\qquad Z_2(r\to \infty) = 1, \qquad f(r\to \infty) = 0
\ee
and
\be
Z_1(r\to 0) = 0,\qquad Z_2(r\to 0) = -1; \qquad f(r\to 0) = \frac{2}{\sqrt{m_0e_0a}}\,\, \sqrt{\left| \frac{m}{n}\right|}
\ee
which agrees with the topological boundary conditions imposed on the field $Z_a$.
The general solution is axially symmetric, and on the $x^3$-axis, corresponding to
$z=0$,  we have
\be
Z_1(0,0,x^3) = 0,\qquad Z_2(0,0,x^3) = e^{i\,n\,\xi}, \qquad
f(0,0,x^3) = \sqrt{\frac{\mid  m \mid}{\mid n\mid} \frac{2(1-\cos\xi) }{m_0e_0a }}
\ee
thus, the solutions are regular everywhere in space.

\begin{figure}[hbt]
\lbfig{f-2}
\begin{center}
\includegraphics[height=.18\textheight, angle =0]{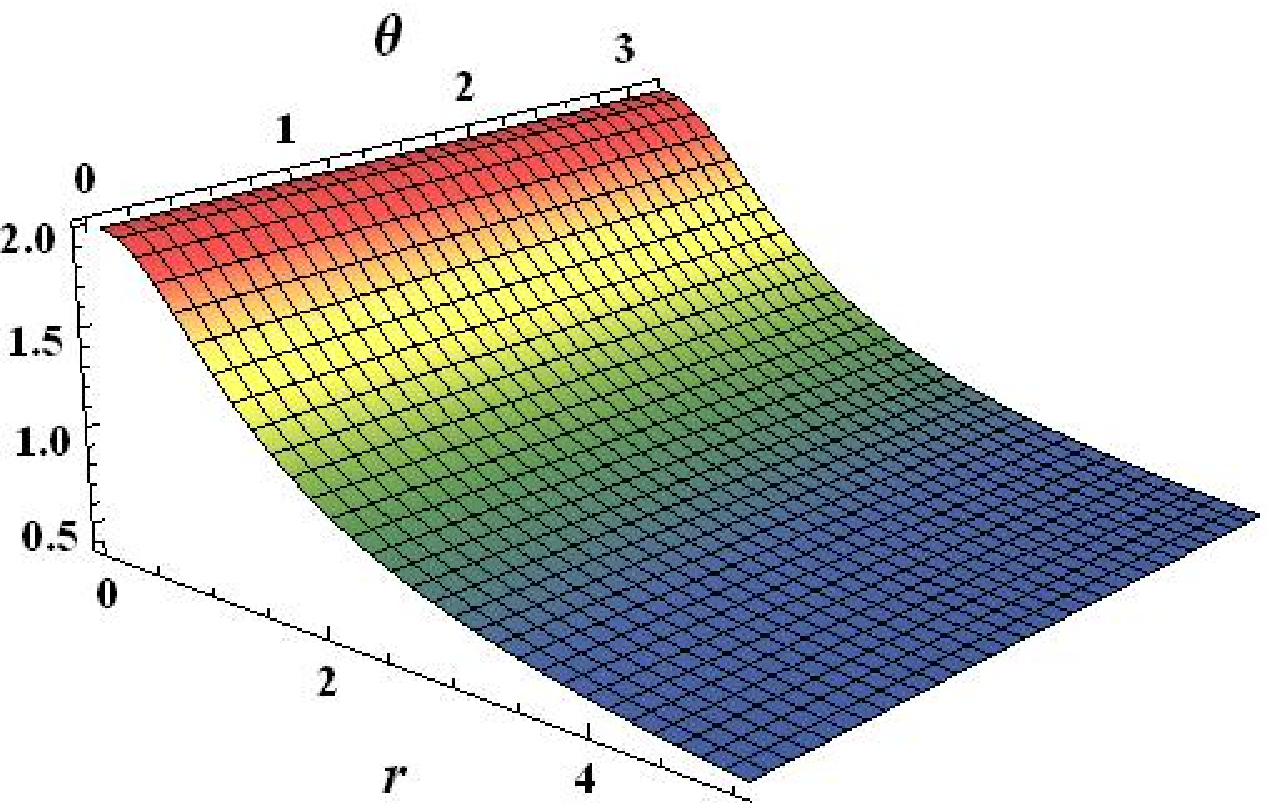}
\includegraphics[height=.18\textheight, angle =0]{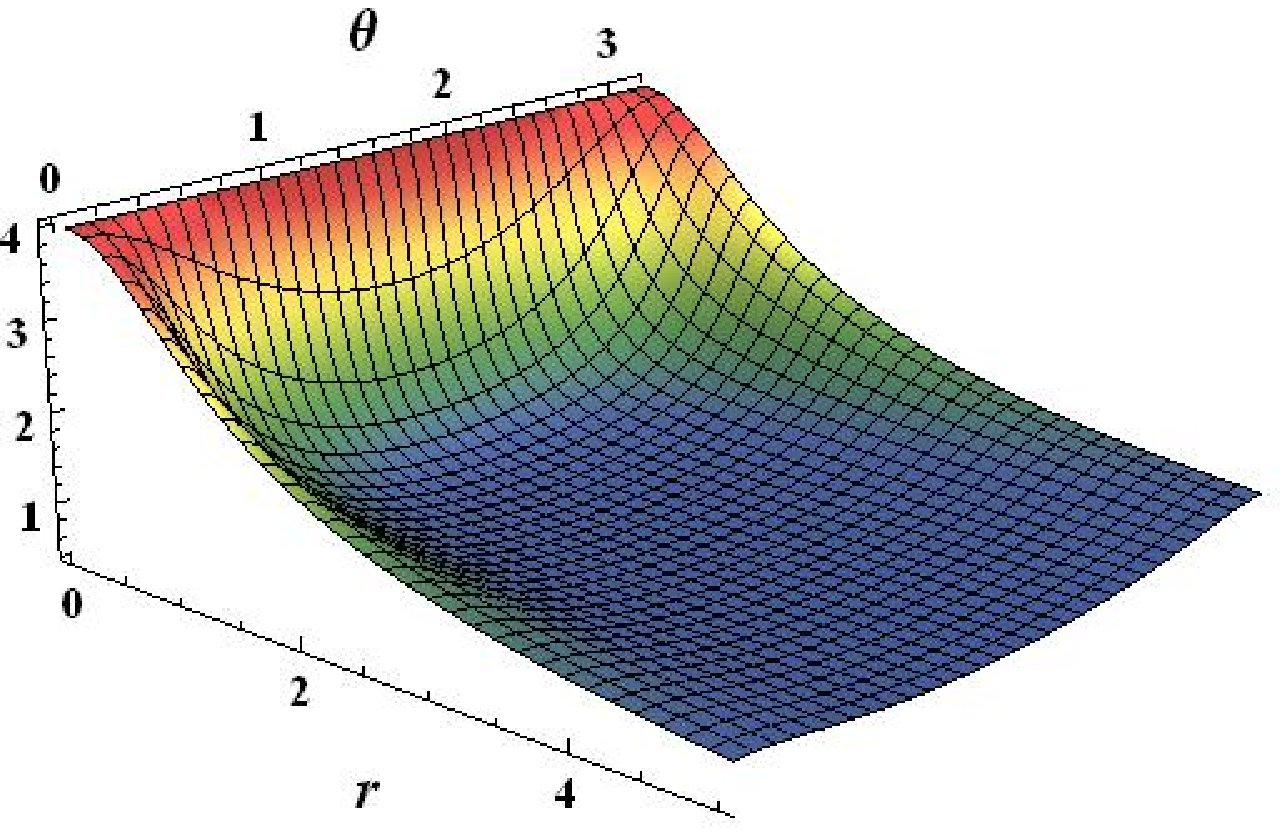}
\includegraphics[height=.19\textheight, angle =0]{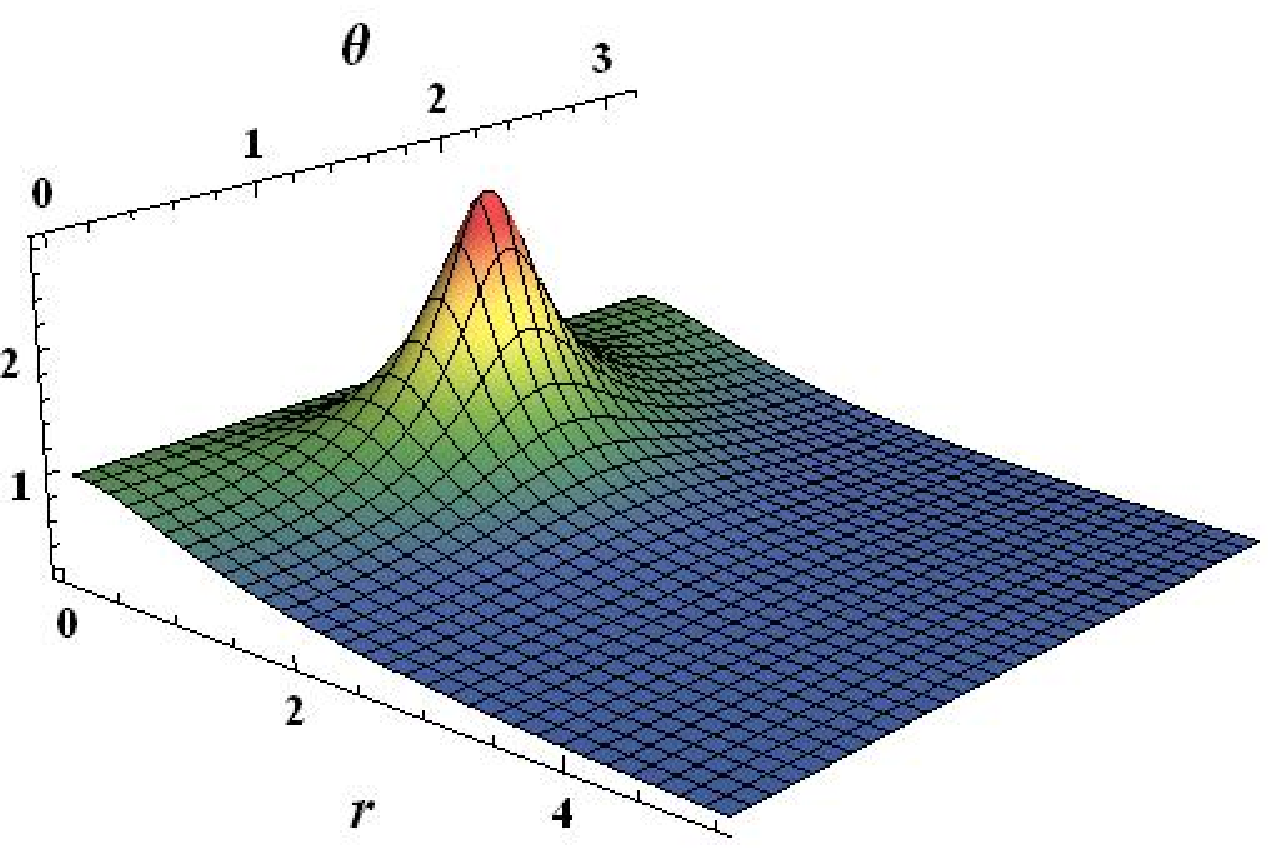}
\end{center}
\caption{\small
The function $f(r,\theta))$ for the solutions \re{solution-fields} of the model  \re{ua845}
 at $a=1$, $m_0=1$, $e_0=1$, for $n=1,m=1$ (left plot),
 $n=1,m=4$ (middle plot), and  $n=4,m=1$ (right plot)
}
\end{figure}

\begin{figure}[hbt]
\lbfig{f-3}
\begin{center}
\includegraphics[height=.40\textheight, angle =-90]{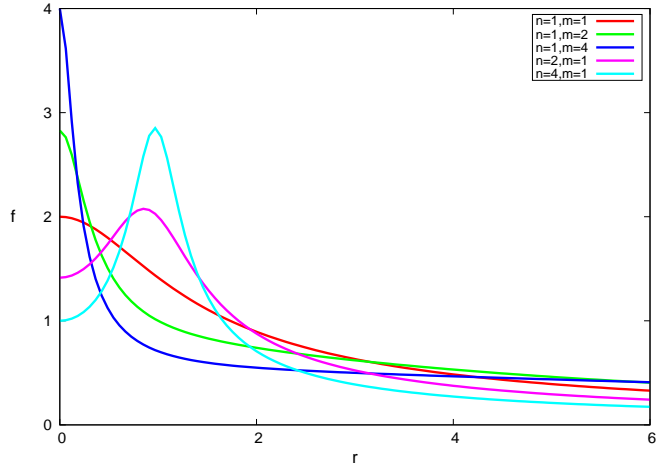}
\end{center}
\caption{\small
The function $f(r,\theta=\pi/2))$ of the self-dual solutions \re{solution-fields} for a few values of the integers  $n,m$, at $a=1$, $m_0=1$, $e_0=1$.
}
\end{figure}

In Figs.~\ref{f-2}-\ref{f-3} we show the function $f$ in terms of spherical coordinates $r,\theta$. For $n=m$ the solutions
possess spherical symmetry. For $n\neq m $, the configuration becomes axially symmetric, it oblate for $n>m$ and it is prolate for
$n<m$.

\begin{figure}[hbt]
\lbfig{f-1}
\begin{center}
\includegraphics[height=.24\textheight, angle =0]{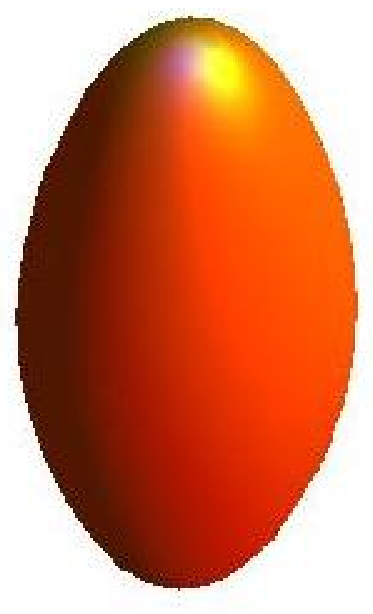}
\includegraphics[height=.23\textheight, angle =0]{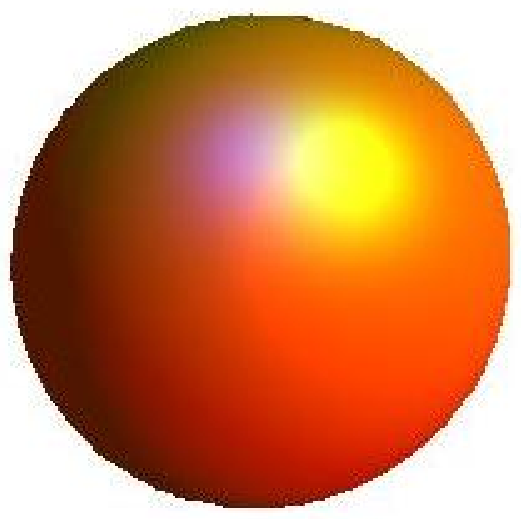}
\includegraphics[height=.19\textheight, angle =0]{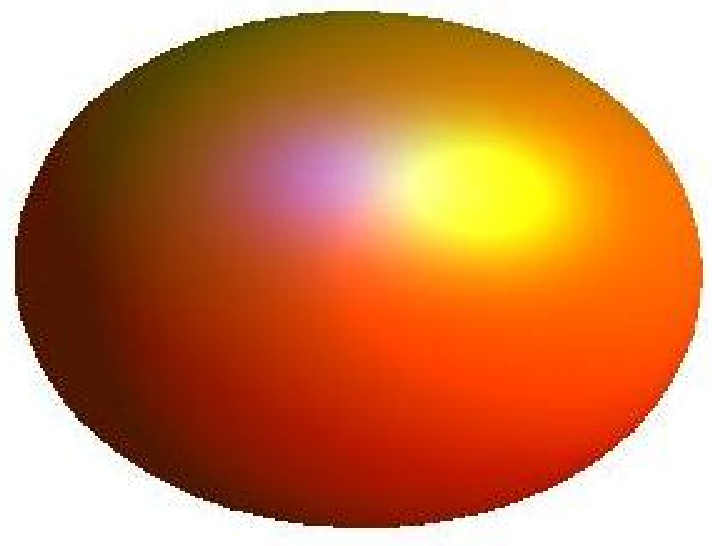}
\end{center}
\caption{\small
The  isosurfaces of the energy density of the $n=1,m=4$ (left plot),
 $n=2,m=2$ (middle plot), and  $n=4,m=1$ (right plot) solutions of the model  \re{ua845}
 at $a=1$, $m_0=1$, $e_0=1$.
}
\end{figure}

The solutions \re{solution-fields} can be written in the spherical coordinates in a more transparent form. Indeed,
using the expressions \re{tor-sphere}, we can write the
energy density of the general solution \re{solution-fields} as
\be
\mathcal{E} = \frac{16 m_0}{e_0\,a^3}|mn |^3 \frac{((r/a)^2+1)}{[4(\rho/a)^2 (m^2-n^2) + n^2 ((r/a)^2+1)^2]^2} \, .
\label{energydensity}
\ee
If $m^2=n^2$, the configuration becomes spherically symmetric, then
\be
\mathcal{E} = \frac{16 m_0}{e_0\,a^3} \frac{ n^2}{((r/a)^2+1)^3}
\ee
Note that in both cases the energy density decays as $1/r^6$ as $r\rightarrow \infty$. In addition, it scales as $1/a^3$, and so, the total energy is scale invariant and that is a consequence of the conformal invariance of the model.

In Fig.~\ref{f-1} we display the energy density iso-surfaces (see \re{energydensity}) for the cases $(n=1,m=4)$,
$(n=2,m=2)$, and  $(n=4,m=1)$.   Note that all these configurations have the same total energy.

Finally, let us  note that the solutions for the $Z_a$ fields given in \re{solution-fields} do not
depend on the arbitrary scale parameter $a$, as one should expect since they scalar under conformal transformations
(see \re{conf-trans}). The function $f$ however scales as $1/\sqrt{a}$, and that is a consequence of the fact it is
not a scalar under dilatations and special conformal transformations, see \re{conf-f}. Thus, similar to the self-dual
soliton solution of the non-linear $O(3)$ sigma model in 2+1 dimensions \cite{Polyakov:1975yp}  the instanton solution
of the Yang-Mills theory
in Euclidian four-dimensional space \cite{Belavin:1975fg}, and the exact Hopfions constructed in \cite{hopfion99}
those field configurations (except for $f$) are  scale invariant.

\section{Conclusions}
The main purpose of this work was to construct exact analytical and regular self-dual solutions of a Skyrme theory with target space $S^3$. The crucial ingredient that made that possible was the conformal symmetry of the self-duality equations in three space dimensions. On its turn, such symmetry was possible due to the fact that the strengths  of the couplings of the quadratic and quartic terms in the action have a space dependence encoded in a  quantity $f$. The physical nature of such quantity is still to be understood, but it is quite natural to relate it to low energy expectation values of fields of a more fundamental theory in higher energies that would contain our Skyrme model as a  low energy effective theory. Note from \re{conf-f} and \re{ababtransf} that the quantity $f$ transforms under the conformal group, in the same way as $\(A_i\,B_i\)^{1/6}$, i.e. a fractional power of the topological charge density. In fact, $f$ and $\(A_i\,B_i\)^{1/6}$ differ by a multiplicative constant, when evaluated on the solutions \re{solution-fields} for the case $m^2=n^2$, i.e the solutions with spherically symmetric energy densities.  Such a fact could perhaps be a hint on how one could try to extend our model by a scalar dilation type field or even vector fields.

Certainly our results open the way for further investigations on the properties of the proposed Skyrme model, and perhaps on its possible physical applications. Of course, it would be interesting to study how the conformal symmetry could be broken leading to scale dependent solutions and bringing a physical scale to the theory. The introduction of a potential or even of the dilation field mentioned above are some of the possibilities. It would also be important to investigate  the rotational modes of the solutions and their semi-classical quantization. Rotating solutions not only would break the conformal symmetry but also would split the energy degeneracies  of our self-dual spectrum. We hope to report on those issues elsewhere.

\vspace{1cm}

\appendix

\section{Toroidal Coordinates}
\label{sec:appendix}

Here we give some useful formulas  related to the toroidal coordinates \re{toroidal}, and that are needed for the explicit calculations leading to the exact solutions \re{solution-fields}. Inverting the relations \re{toroidal} one gets that
\be
z=\frac{4\,a^2\(x_1^2+x_2^2\)}{\(x_1^2+x_2^2+x_3^2+a^2\)^2}\;;\quad
\xi={\rm ArcTan}\left[\frac{2\,a\,x_3}{\( x_1^2+x_2^2+x_3^2-a^2\)} \right];\quad
\varphi={\rm ArcTan}\( \frac{x_2}{x_1}\)
\label{tor-sphere}
\ee
The metric in toroidal coordinates is $ds^2= h_z^2\,dz^2+h_{\xi}^2\,d\xi^2+h_{\varphi}^2\,d\varphi^2$, with scaling factors being
\be
h_{z}=\frac{a}{p}\,\frac{1}{2\,\sqrt{z\(1-z\)}}\qquad\quad
h_{\xi}=\frac{a}{p}\,\sqrt{1-z}\qquad\quad
h_{\varphi}=\frac{a}{p}\,\sqrt{z}
\label{scalingfactors}
\ee
The volume element is then
\be
dx^1\, dx^2\,dx^3=\frac{1}{2}\,\frac{a^3}{p^3}\, dz\,d\xi\,d\varphi
\ee
Note that
\be
r^2=x_1^2+x_2^2+x_3^2=a^2\,\frac{\(1+\sqrt{1-z}\, \cos \xi\)}{\(1-\sqrt{1-z}\, \cos \xi\)}
\label{r2}
\ee
Therefore the spatial infinity corresponds to $z=0$, and $\xi=0$ (or $2\,\pi$). The $x^3$-axis corresponds to $z=0$,   for $0 <\xi <2\,\pi$.  The origin corresponds to $z=0$, and $\xi=\pi$.
In addition, $z=1$ corresponds to the circle $x_1^2+x_2^2=a^2$, and  $x_3=0$.

The unit vectors are defined as ${\vec e}_{\zeta}= \frac{1}{h_{\zeta}}\,\frac{d\,{\vec r}}{d\,\zeta}$, for $\zeta\equiv z\,,\, \xi\,,\, \vp$, and so we have that
\br
{\vec e}_{z}&=&\frac{1}{p}\left[(\sqrt{1 - z} - \cos\xi) \cos\vp\,{\vec e}_1+ (\sqrt{1 - z} - \cos\xi) \sin
  \vp\,{\vec e}_2 -\sqrt{z} \sin\xi\, {\vec e}_3\right]
  \nonumber\\
  {\vec e}_{\xi}&=&-\frac{1}{p}\left[
  \sqrt{z} \cos\vp \sin\xi\,{\vec e}_1+ \sqrt{z} \sin\vp \sin\xi\,{\vec e}_2+ (\sqrt{1 - z} - \cos\xi)\,\,{\vec e}_3  \right]
 \label{unitvectors}\\
  {\vec e}_{\vp}&=& -\sin\vp\, {\vec e}_1+\cos\vp\,{\vec e}_2
   \nonumber
\er
where ${\vec e}_i$, $i=1,2,3$, are the unit vectors in Cartesian coordinates.

\vspace{1cm}

\noindent{\bf Acknowledgements:} The authors are grateful to Profs. Wojtek Zakrzewski and Nobuyuki Sawado
for valuable discussions. LAF is partially supported by CNPq-Brazil. YS
thanks Ilya Perapechka for relevant discussion. YS is grateful to
Funda\c c\~ao de Apoio \`a Pesquisa do Estado de S\~ao Paulo, FAPESP, for the financial support under
the grant 2015/25779-6, he also gratefully
acknowledges support from the Russian Foundation for Basic Research
(Grant No. 16-52-12012), the Ministry of Education and Science
of Russian Federation, project No 3.1386.2017, and DFG (Grant LE 838/12-2).
YS would like to thank the
Instituto de F\'{i}sica de S\~{a}o Carlos for its kind
hospitality.

\vspace{1cm}

\begin{small}

\end{small}
\end{document}